\documentclass[preprint2]{aastex}

 \shorttitle{Time Delay} \shortauthors {Fargion}

\newcommand{\ffrac}[2]  {\left( \frac{#1}{#2} \right)}

\begin{document}

\title{\bf Time Delay Between Gravitational Waves and Neutrino Burst From a
Supernova Explosion: a Test  for the Neutrino Mass }

\author{D. Fargion }
\affil{ Physics Department, Rome University 1 \\ ''La
Sapienza'',INFN,  P.le A. Moro 2, 00185 Roma, Italy.}

\email{Daniele.Fargion@roma1.infn.it}



\begin{abstract} \leftskip = 0cm \rightskip = 0cm
Published in LETTERE   AL   NUOVO   CIMENTO $ VOL.   31, N. 15,
p.~499-500;$ \newline \centering \bf{8 August 1981}. 
\centering{\bf{}}\vspace{1cm}
\end{abstract}

The gravitational collapse of a massive star is one of the most
powerful mechanisms of simultaneous emission of neutrinos and
gravitational waves. Almost all the present models for a
supernova explosion predict a typical outburst of $10^{53}$ $erg$
in its neutrino flow (namely electronic anti-neutrinos or neutrino
pairs) with an average energy of 15 MeV for each neutrino. Most
of these models predict that a relevant fraction of the neutrinos
should be emitted by the supernova on a time scale of 0.1 s
\footnote{J. SHRAMM: D. N. Proceedings of Neutrino  '79 (Bergen,
1979), p. 503} or even in shorter pulses of $10^{-3} s$
\footnote{C. CASTAGNOLI, P.  GALEOTTI and O. SAAVERDRA: Astophys.
Space Sci., \bf{55} , 511 (1978)}. The gravitational waves
emitted in stellar collapse have the same characteristic time
scale, since both processes are due to the same physical
mechanism, i.e. a time of $10^{-3} \div 10^{-1}s$. Therefore, two
signals, of different nature, originated within 0.1 s from the
super nova explosion will be radiated in the outer space. If the
neutrinos (as well as the gravitons) are massless particles, than
the original interval (it any) between the two signals will be
freezed forever. On the contrary, if the neutrinos have a mass,
the time delay will increase during the propagation because of
the slower velocity of the neutrinos with respect to the massless
gravitons: namely for a neutrino mass  $m_{\nu} = 15 \eta eV$
($\eta \sim 1$) , with an emission energy $E_{\nu}= 15 MeV$, and
for a source (the supernova) located in our Galaxy at a distance
$L = 10 kpc$ from us, the time delay between the two signals
detected on the Earth will increase by a factor

\begin{equation}
\Delta \tau \simeq \frac{L}{2c} \ffrac{m_{\nu}}{E_{\nu}}^2 = 0.5
\; \eta^2 \; s
\end{equation}

For extra-galactic sources  at $L= 10^{2} \div 10^{3}$ kpc the
delay will be $\Delta \tau \simeq (5 \div 50 )\; \eta^2 \; s$.
Since the sensibility of the present neutrino detectors and
gravitational antennae permit us to observe supernova events
within our Galaxy (and in the near future even within near group
of galaxies) \footnote{E. AMALDI and G. PIZZELLA: Astrofisica e
cosmologia, gravitazione, quanti e relatività (Firenze, 1979), p.
283 },\footnote{J. LEARNED and D. EICHLER: Scientific American
(February 1981), p. 104. C)}
 since their time accuracy is roughly
$10^{-3} s$, the time delay measure is possible and it could give
an estimate of the neutrino mass. No appreciable time delay will
infer strong limits to the neutrino mass. Analogous arguments for
a time correlation between optical events $^4$ and neutrino burst
or neutrino-neutrino bursts \footnote{N. CABIBBO: Astrophysics and
Elementary Particles, Common Problems (Roma,  1980), p.  229.} at
different energies (originated by a supernova explosion) seem to
be, at the moment, less praticable or less conclusive. The author
wishes to thank Dr. C. COSMELLI, Prof. N. CABIBBO and Prof. J.
LINSLEY for stimulating conversation and Prof.  E.  AMALDI  for
kind encouragement.

\end{document}